\documentclass[]{aastex61}

\def\Ha{H$\alpha$}
\def\Hb{H$\beta$}
\def\hii{{\sc H\thinspace ii}}

\def\oii{[{\sc O\thinspace ii}]}
\def\oiii{[{\sc O\thinspace iii}]}
\def\msol{\rm\,M_\odot}

\def\kms{{\rm\,km\,s^{-1}}}
\def\ergs{{\rm\,erg\,s^{-1}}}
\def\cc{{\rm\,cm^{-3}}}

\def\mrka{Mrk~71-A}

\received{2017 September 12}
\accepted{2017 October 8}

\shorttitle{CO in \mrka:  Superwind Suppressed}
\shortauthors{Oey et al.}

\begin{document}

\title{Dense CO in Mrk 71-A:  Superwind Suppressed in a Young
  Super Star Cluster} 

\correspondingauthor{M. S. Oey}
\email{msoey@umich.edu}

\author[0000-0002-5808-1320]{M. S. Oey}
\affil{Department of Astronomy \\
University of Michigan \\
1085 South University Ave. \\
Ann Arbor, MI   48109-1107, USA}

\author{C. N. Herrera}
\affiliation{Institut de Radioastronomie Millim\'etrique \\
  300 Rue de la Piscine\\
  Domaine Universitaire\\
  38406, Saint-Martin-d'H\`eres, France}

\author{Sergiy Silich}
\affiliation{Insituto Nacional de Astrof\'isica, Optica y Electr\'onica\\
  AP 51\\
  72000 Puebla, Mexico}

\author{Megan Reiter}
\affiliation{Department of Astronomy \\
University of Michigan \\
1085 South University Ave. \\
Ann Arbor, MI   48109-1107, USA}

\author{Bethan L. James}
\affiliation{Space Telescope Science Institute \\
  3700 San Martin Drive \\
  Baltimore, MD   21218, USA}

\author{A. E. Jaskot}
\altaffiliation{Hubble Fellow}
\affiliation{Department of Astronomy \\
  University of Massachusetts \\
  Amherst, MA   01003, USA}

\author{Genoveva Micheva}
\affiliation{Department of Astronomy \\
University of Michigan \\
1085 South University Ave. \\
Ann Arbor, MI   48109-1107, USA}

\begin{abstract}

We report the detection of CO($J=2-1$) coincident with the super star
cluster (SSC) \mrka\ in the nearby Green Pea analog galaxy, NGC~2366.  Our NOEMA
observations reveal a compact, $\sim 7$ pc, molecular cloud whose mass
($10^5 \msol$) is similar to that of the SSC, consistent with a  high 
star-formation efficiency, on the order of 0.5.  There are 
two, spatially distinct components separated by 11 $\kms$.  If
expanding, these could be due to momentum-driven, stellar wind
feedback.  Alternatively, we may be seeing the remnant infalling, colliding
clouds responsible for triggering the SSC formation.  The 
kinematics are also consistent with a virialized system.
These extreme, high-density, star-forming conditions 
inhibit energy-driven feedback; the
co-spatial existence of a massive, molecular cloud with the SSC
supports this scenario, and we quantitatively confirm that any
wind-driven feedback in \mrka\ is momentum-driven, rather than
energy-driven.  Since \mrka\ is a candidate Lyman continuum emitter,
this implies that energy-driven superwinds may not be a necessary
condition for the escape of ionizing radiation.
In addition, the detection of the nebular continuum emission yields
an accurate astrometric position for the \mrka.  
We also detect four other massive, molecular clouds in this
giant star-forming complex.

\end{abstract}


\section{Introduction} \label{sec:intro}

How and when do newborn super star clusters (SSCs), like the progenitors of
globular clusters, emerge from dense gas and clear their surroundings?   
The classical model for starburst mechanical feedback holds that
stellar winds and supernovae inside SSCs collide and merge to form
strong superwinds \citep{chevalier85}.  This hot ($\gtrsim10^7$ K),
shock-heated gas carries off most of the 
newly synthesized heavy elements produced by the massive stars and
their supernovae.  In dwarf galaxies, these metals could be ejected
into the intergalactic medium, retarding chemical evolution of the
host galaxy.  However, more recent developments in our
understanding of both massive star feedback and chemical abundance
patterns suggest that this picture is more complex.  Theoretical
work suggests that catastrophic cooling and large ambient gas pressure
may inhibit the development of superwinds and wind-driven superbubbles
\citep{silich07, silich17}, 
and radiation feedback is suggested to be important in the youngest
and densest clusters \citep[e.g.,][]{freyer03,krumholz09}.
On the observational side, multiple stellar populations with
differing abundances are now ubiquitous in individual globular
clusters \citep[e.g.,][]{gratton12}, suggesting that star formation
may have recurred in these objects.  So then what factors dominate the
interaction between a brand-new SSC and its gaseous environment?
Observations of newly born SSCs that are still clearing their gas are
vital to revealing these processes.  Here, we present remarkable CO
observations revealing that a young, $10^5 \msol$ SSC, \mrka, located
in a relatively clear environment, nevertheless 
coexists with an extremely compact, molecular cloud of comparable mass.

\begin{figure}[ht!]
\plotone{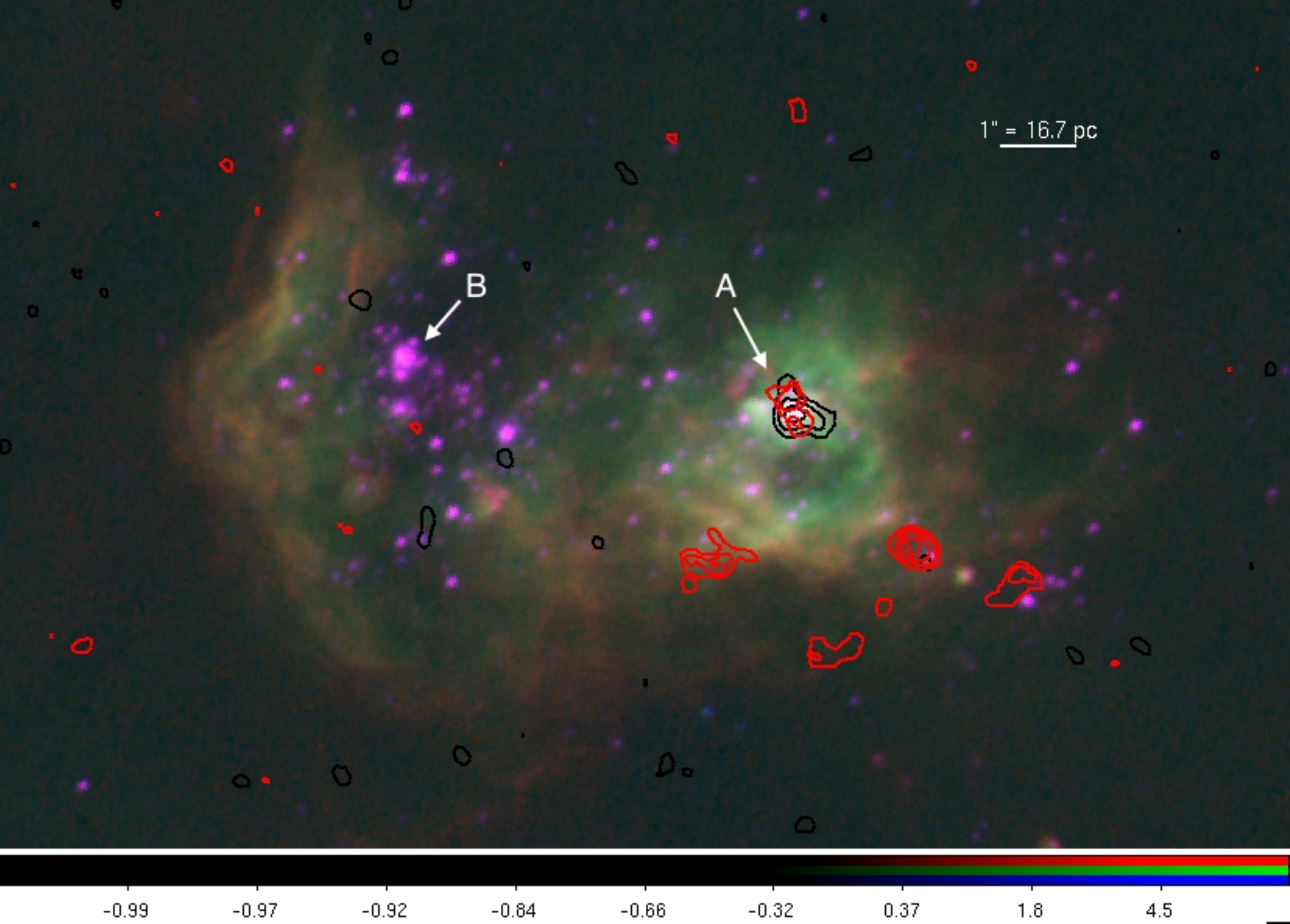}
\caption{Composite image of Mrk~71 using {\sl HST}/WFC3
  data from \citet{james16}.  Red, green, blue correspond to 
\oii\ $\lambda3727$ (F373N), \oiii $\lambda5007$ (F502N), and
blue continuum (F438W), respectively.  Red contours show line
emission of CO(2--1) detected with NOEMA, integrated over 62 --
90.6 $\kms$, selecting pixels with emission $>2\sigma$.  Contour
levels are 0.03, 0.06, 0.09, 0.12, and 0.15 Jy/beam km/s.  Black
contours show the corresponding continuum emission integrated over 3.6
GHz, excluding 10 MHz of line emission,
with contour levels at 2, 3, and $4\times 10^{-4}$ Jy/beam.
North is up, east to the left.
\label{f_HSTCO}}
\end{figure}

Mrk 71 is a well-studied starburst complex in the nearby barred Im
galaxy, NGC 2366, at a distance of 3.4 Mpc \citep{tolstoy95}.  It hosts two
SSCs, Mrk 71-A and B (Figure~\ref{f_HSTCO}).  Mrk 71-B is an exposed
cluster with mass $1.5\times 10^4 \msol$ \citep{micheva17} and
age of $3-5$ Myr constrained by the presence of classical WR stars
\citep{D00}.  \mrka, however, strongly dominates the complex, producing about 
ten times more ionizing photons, implying an SSC stellar mass of
$M_\star=1.4\times 10^5 \msol$ \citep{micheva17, D00}.  To date, \mrka\ has only been
detected as a compact, extreme excitation \hii\ region.  No stellar
features are spectroscopically detected, but strong nebular continuum
is observed, including inverse Balmer and Paschen jumps 
\citep[e.g.,][]{GD94,guseva06}.  It
is marginally spatially resolved with the {\sl Hubble Space Telescope}
(HST), implying a radius of $\lesssim 1$ pc.  Thus \mrka\ has an
extremely high nebular surface brightness and is remarkably similar to
a massive, ultracompact \hii\ region \citep{D00}, but
hosting a $10^5 \msol$ newly-formed, embedded cluster.  

The Mrk 71/NGC 2366 system is especially interesting owing to its status
as a local analog to extreme ``Green Pea'' galaxies.  These are objects at
$z\sim0.2$ \citep{cardamone09} that represent a
compelling class of local, candidate Lyman continuum-emitting galaxies.
They are low-mass, compact starbursts characterized by high ratios of
\oiii$\lambda$5007/\oii$\lambda3727\gtrsim 5$ and \Hb\ equivalent
widths $> 100$ \AA, and they are in many ways good analogs of
high-redshift starbursts.  We showed that among these objects, those
with the most extreme ionization parameters are strong candidate Lyman
continuum emitters \citep[LCEs;][]{JO13}.  Our prediction was
dramatically confirmed when \citet[][b]{izotov16a} detected
Lyman continuum emission from all five of their targeted Green Peas, thereby
instantly doubling the number of confirmed local LCEs.  The properties
of Mrk~71 are quantitatively consistent with those of Green Peas and,
like those objects, Mrk~71 shows a variety of features that are
consistent with optically thin Lyman continuum
\citep{micheva17}.  Thus, Mrk 71/NGC 2366 may also offer a unique opportunity
to study what properties facilitate the escape of ionizing radiation.

\section{CO Observations}

We obtained observations of Mrk 71 on 2016 December 14 and 15, with the
Northern Extended Millimeter Array (NOEMA) at Plateau de Bure, in the
CO ($J=2-1$) line at 230.538~GHz, with a local standard of rest velocity of 90~$\kms$.
The observations were carried out with 8 antennae in Configuration A, with 
baselines between 45~m and 760~m.  We used the Widex correlator, with total 
bandwidth of 3.6~GHz and native spectral resolution 1.95~MHz (2.6~$\kms$).
The quasars 0716+714 and 0836+710 were observed as phase and amplitude calibrators, 
3C~84 as bandpass calibrator, and the stars MWC~349 and LkHA~101 (1.91~Jy and 0.56~Jy 
at 230.5~GHz) as absolute flux calibrators.  System temperatures were between 90 and 240 K, 
and the precipitable water vapor $\sim$2--3~mm. 
Data reduction, calibration and imaging were performed with CLIC and MAPPING 
software of GILDAS\footnote{http://www.iram.fr/IRAMFR/GILDAS/}, using standard 
procedures.  Images were reconstructed using natural weighting, resulting
in a synthesized beam of 0$\farcs$48$\times$0$\farcs$35 (PA=55.8$^{\circ}$) and 
rms noise in the CO map of 2.2 mJy beam$^{-1}$ in a 2.6~$\kms$
channel.  The continuum map has rms noise of 75 $\mu$Jy beam$^{-1}$.

\begin{deluxetable*}{ccccccccc}
\tablecaption{CO(2--1) Observations of Mrk 71 \label{t_mrk71a}}
\tablewidth{0pt}
\tablehead{
\colhead{Cloud\tablenotemark{a}} & {Size} &
\colhead{$v_{\rm LSR}$} &
  \colhead{$\Delta v$\  \tablenotemark{b}} &
  \colhead{Peak}  &
  \colhead{Flux}  &
  \colhead{$I_{\rm CO}$} &
    \colhead{Flux mass\tablenotemark{c}} &
    \colhead{Virial mass\tablenotemark{d}} \\
\colhead{} & \colhead{pc} &
\colhead{[km/s]} &
\colhead{[km/s]} &
\colhead{[mJy]} &
\colhead{[mJy km/s]} &
\colhead{[K km/s]} &
\colhead{[$10^4\ \rm M_{\odot}$]}  &
\colhead{[$10^4\ \rm M_{\odot}$]}
}
\startdata
1 & 9.0 & 77.7$\pm$0.7 & 15.9$\pm$1.7 & 10.7 $\pm$1.0 &
181$\pm$25 &13$\pm$2 & 14$\pm$2 & 22$\pm$5 \\
1 ($>$3$\sigma$) & 5.5 & 78.1$\pm$0.7 & 16.1$\pm$1.7 & 7.7$\pm$0.7  &
131$\pm$18& 15$\pm$2 & 10$\pm$1 & 14$\pm$3 \\
1-blue & 10.7 &73.6$\pm$0.7 & 6.3$\pm$1.5 & 12.0$\pm$2.0 &
80$\pm$24 & 5.9$\pm1.7$  & 6.2$\pm$1.8 & 4.0$\pm$2.0 \\
1-red & 5.4 & 82.6$\pm$1.2 & 9.0$\pm$2.8 & 9.3$\pm$1.4 &
89$\pm$31 & 6.6$\pm$2.3 &  6.9$\pm$2.4 & 4.2$\pm$2.6 \\
2 & 8.5 & 78.7$\pm$0.8 & 15.3$\pm$2.0 & 13.1$\pm$1.4 &
213$\pm$36 & 12$\pm$2 & 16$\pm$3 & 19$\pm$5 \\
3 & 9.7 & 75.0$\pm$0.4 & 8.4$\pm$1.0 & 20.1$\pm$2.0 &
180$\pm$28 & 14$\pm$3 & 14$\pm$2 & 5.7$\pm$1.3 \\
4 & 8.9 & 79.1$\pm$0.4 & 7.6$\pm$1.0 & 17.0$\pm$2.0 &
137$\pm$25 & 9$\pm$2 & 11$\pm$2 & 4.6$\pm$1.3 \\
5 & 8.6 & 73.5$\pm$0.4 & 6.1$\pm$0.9 & 15.6$\pm$1.9 &
101$\pm$19 & 7$\pm$3 & 7.8$\pm$1.5 & 3.0$\pm$0.9 \\
\enddata
\tablenotetext{a}{Values measured from pixels with fluxes $>2\sigma$, unless
  otherwise specified.  ``Blue'' and ``red'' denote the 
  components of Cloud 1.}
\tablenotetext{b}{FWHM of CO line.}
\tablenotetext{c}{Mass based on CO intensity estimated using $X_{\rm
  CO} = 50\times10^{20}\rm  cm^{-2}\ (K\ km/s)^{-1}$.  The errors
  shown correspond to measurement uncertainty only, while $X_{\rm CO}$ is
  uncertain to a factor of a few.}
\tablenotetext{d}{Virial mass calculated as $190\ r{\rm (pc)} \times[\Delta
    v \rm (km/s)]^2$  \citep{maclaren88}, where the cloud radius $r$ is
  half the value in Column 2.  Errors shown are estimated only
  from the error in the fit of $\Delta v$.}
\end{deluxetable*}

The 22$\arcsec$ field of view encompasses the entire Mrk~71 complex.
Figure~\ref{f_HSTCO} shows a 3-color image using {\sl HST}/WFC3 data
in \oii\ $\lambda3727$ (F373N), \oiii $\lambda5007$ (F502N), and
blue continuum (F438W) superposed with red contours showing the 
continuum emission at 231~GHz within a bandwidth of 3.6~GHz, and black
contours showing line emission of CO(2--1).  Five sources
are detected, one of which we determine coincides with \mrka, as
follows.  The {\sl HST} absolute astrometry has uncertainty 
of $\sim 1\arcsec$, while the NOEMA 
astrometry is good to 0.05$\arcsec$.  However, since \mrka\ is a
strong nebular continuum source, our detected mm source
must correspond to this target.  Flux densities for \mrka\ from the
VLA reported at 3.6, 6, and 20 cm by Chomiuk \& Wilcots (2009) with
angular resolution of 3.7$\arcsec$ yield a spectral index of --0.13,
indicating a thermal \hii\ region continuum.
Our NOEMA continuum observation of 0.71 mJy is less than the predicted
value of 3.35 mJy extrapolated from the VLA data.  However, we only
detect the high surface-brightness core of \mrka\ in the NOEMA
continuum (Figure~\ref{f_HSTCO}).  Since the VLA beam diameter is about
10$\times$ larger than for our observations, the extended nebular emission
dominates the flux in the large aperture.
The non-detection of any other NOEMA continuum source within the VLA beam
and the agreement of the flux densities with a thermal power-law index
therefore indicate that the observed emission must be due to the same source, \mrka.  In
Figure~\ref{f_HSTCO}, we have therefore aligned the data so that the mm
continuum peak necessarily coincides with the optical \hii\ region centroid.
The position for \mrka\ is thus:
$07^h\ 28^m\ 42^s.716,\ +69^\circ\ 11\arcmin\ 22\arcsec.07$ (J2000) with
uncertainty of 0.05$\arcsec$. 

Figure~\ref{f_HSTCO} shows a detection of CO(2--1)
coincident with the thermal continuum source.  The first two 
lines of Table~1 show the source parameters based on calculating the
flux above $2\sigma$ and $3\sigma$.
Mrk~71 has $12+\log(\rm O/H)=7.89$ (Izotov et al. 1997).
For this low, SMC-like value, the transition ratio  $R_{\rm CO}\equiv$
CO(2--1)/CO(1--0) is $\gtrsim 1$, observed in SMC \hii\ regions (Rubio
et al. 1993), and similar ratios are seen in 30 Doradus 
($R_{\rm CO}=0.95\pm0.06$; Sorai et al. 2001).  
Thus, we take $R_{\rm CO} = 1$.
The CO-to-H$_2$ conversion factor from CO(1--0), $X_{\rm CO}$, is quite
uncertain at this low metallicity;
based on studies of other blue compact dwarfs
\citep{amorin16, bolatto13, leroy11}, we adopt $X_{\rm CO}=50\times
10^{20}\ \rm cm^{-2}/(K \kms)$, yielding a total molecular gas mass of
$M_g=1 \times 10^5\ \msol$.  This cloud mass is similar to our estimated mass
of $M_\star = 1.3\times 10^5\ \msol$ for the enshrouded SSC, based on the
\Ha\ luminosity \citep{micheva17}.  Taken together, the masses 
agree well with the virial mass (Table~1).

However, we caution that $X_{\rm CO}$ is uncertain to a factor of a
few.  If the system is not virial,
the cloud mass may be a substantial upper limit.  On the other hand,
the \mrka\ system strongly resembles the molecular cloud NGC~5253-D1,
which is associated with a radio supernebula and SSC \citep{turner15,
  turner17}.  \citet{turner17} suggest that the SSC may be
overluminous for its mass; this may also apply to
\mrka, where we indeed suggested possible evidence for very massive
stars (VMS) of $>100\ \msol$\ \citep{micheva17}.  If these dominate the SSC luminosity,
then its stellar mass would be overestimated.  In any case, the relative masses
of the cloud and SSC are consistent with an extremely high
star-formation efficiency (SFE), on the order of 0.5.  This is again similar
to NGC~5253-D1 \citep{turner15} and SGMC 4/5-B1 in the Antennae \citep{herrera17}.  

\begin{figure}
\vspace*{-0.5in}
\epsscale{1.0}
\plotone{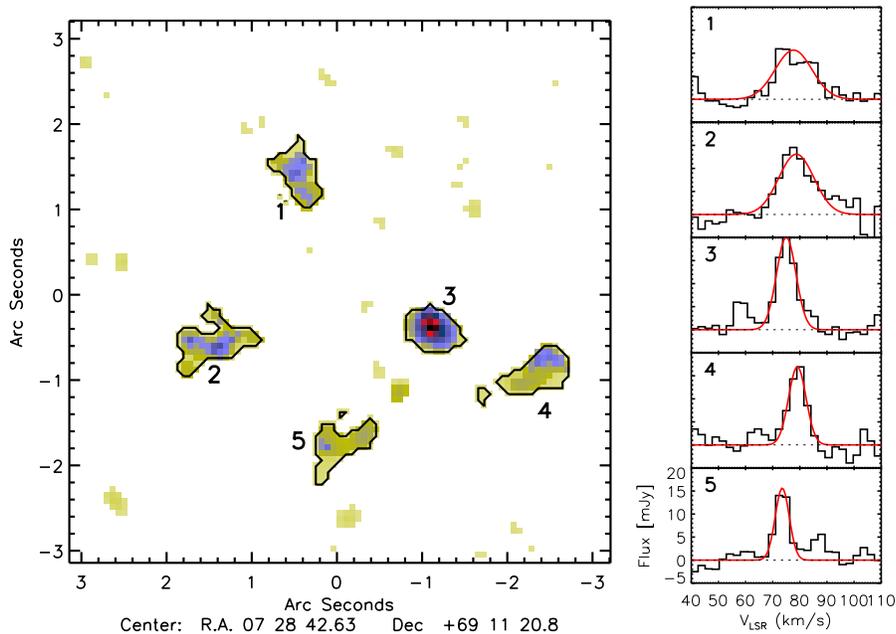}
\vspace*{-3.5in}
  \caption{
Intensity map and CO(2--1) line profiles for all newly detected CO
clouds shown in Figure~\ref{f_HSTCO}.  Source 1 is \mrka.
\label{f_otherclouds}}
\end{figure}

We detect four other sources at $>2\sigma$.  All are in the region
south of Knot A, within 50 pc of the SSC (Figures~\ref{f_HSTCO},
\ref{f_otherclouds}), with no continuum detections.  These are all
molecular clouds similar in mass to Knot A, with parameters
given in Table~1.  Source~2 has a large line width, similar to Source~1
(Figure~\ref{f_otherclouds}).  Source 3 has the highest peak intensity,
but shows no evidence of star formation.

\section{Feedback in Mrk 71-A}

Young SSCs are thought to clear their environment through massive star feedback, but
the details of how and when this takes place are poorly understood.
Alternatively, extremely high SFE may itself account for some
gas clearing \citep{kruijssen12}.  Typically, feedback is
characterized by a model in which the mechanical energy of stellar
winds and supernova ejecta is thermalized, leading to a high central
overpressure.  The hot, shocked gas flows away from the cluster
as a superwind, colliding with the ambient ISM and forming a bubble
that grows in an energy-conserving mode \citep{weaver77,maclow88}.
Through photoionization and radiation pressure, Lyman continuum (LyC)
photons from the hot, massive stars also generate feedback effects that 
can be of comparable magnitude to mechanical feedback, especially for the
youngest and most compact SSCs \citep[e.g.,][but see Silich \&
Tenorio-Tagle 2013; Mart\'inez-Gonz\'alez et
al. 2014]{freyer03,krumholz09,dale15}.  It is essential to understand
how these processes interact to determine the duration of the clearing
process, and how the gas properties and conditions evolve.
\mrka\ offers a unique opportunity to test feedback theories.  Its
mass, compactness, and extreme youth place it in the critical regime
where the dominant processes are unclear, and its proximity, with an
unobscured line of sight, allows high quality estimates of the physical
parameters for quantitative modeling.

\begin{figure}
\epsscale{0.6}
\hspace*{-0.4in}
\plotone{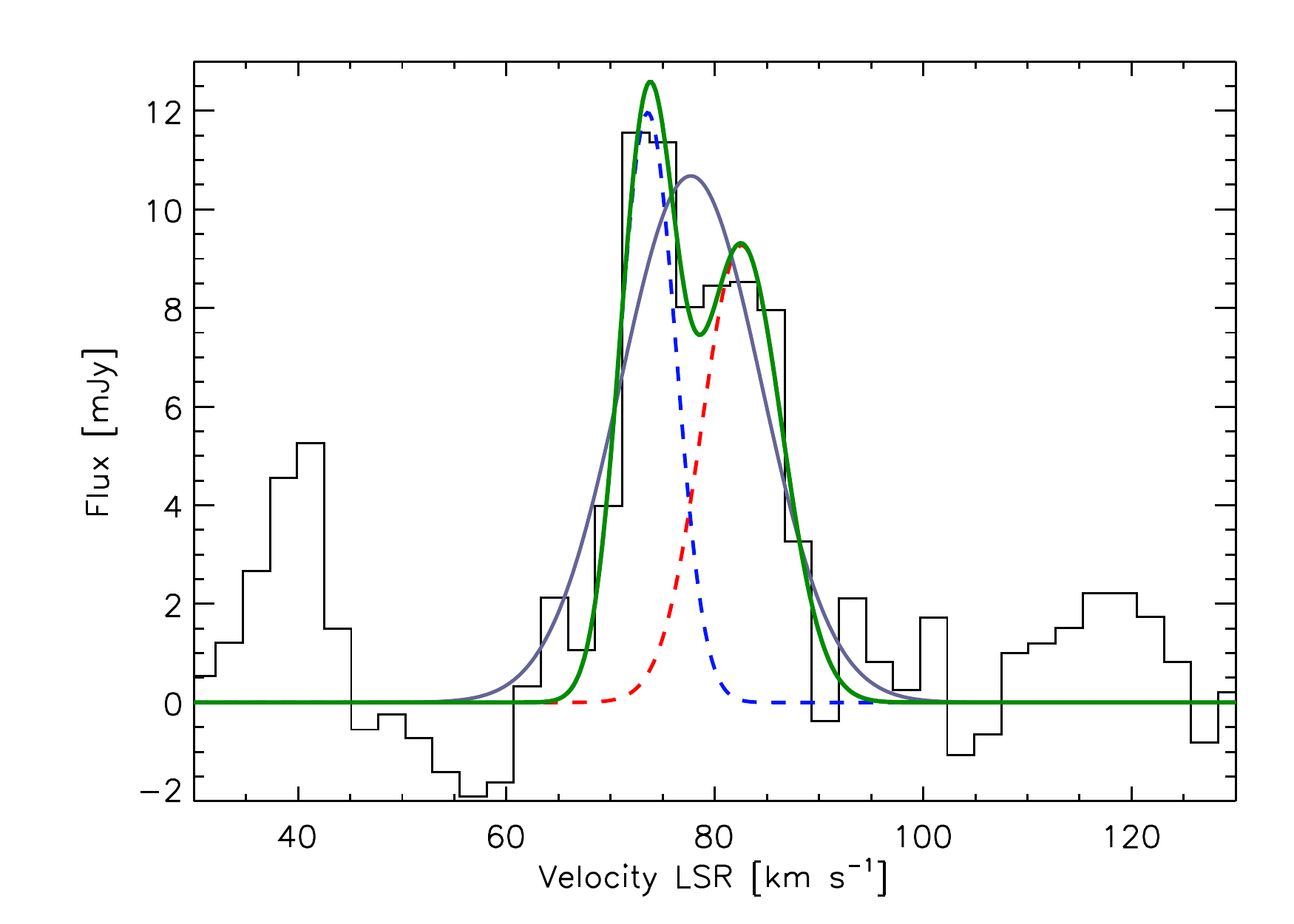}
\hspace*{-0.2in}
\plotone{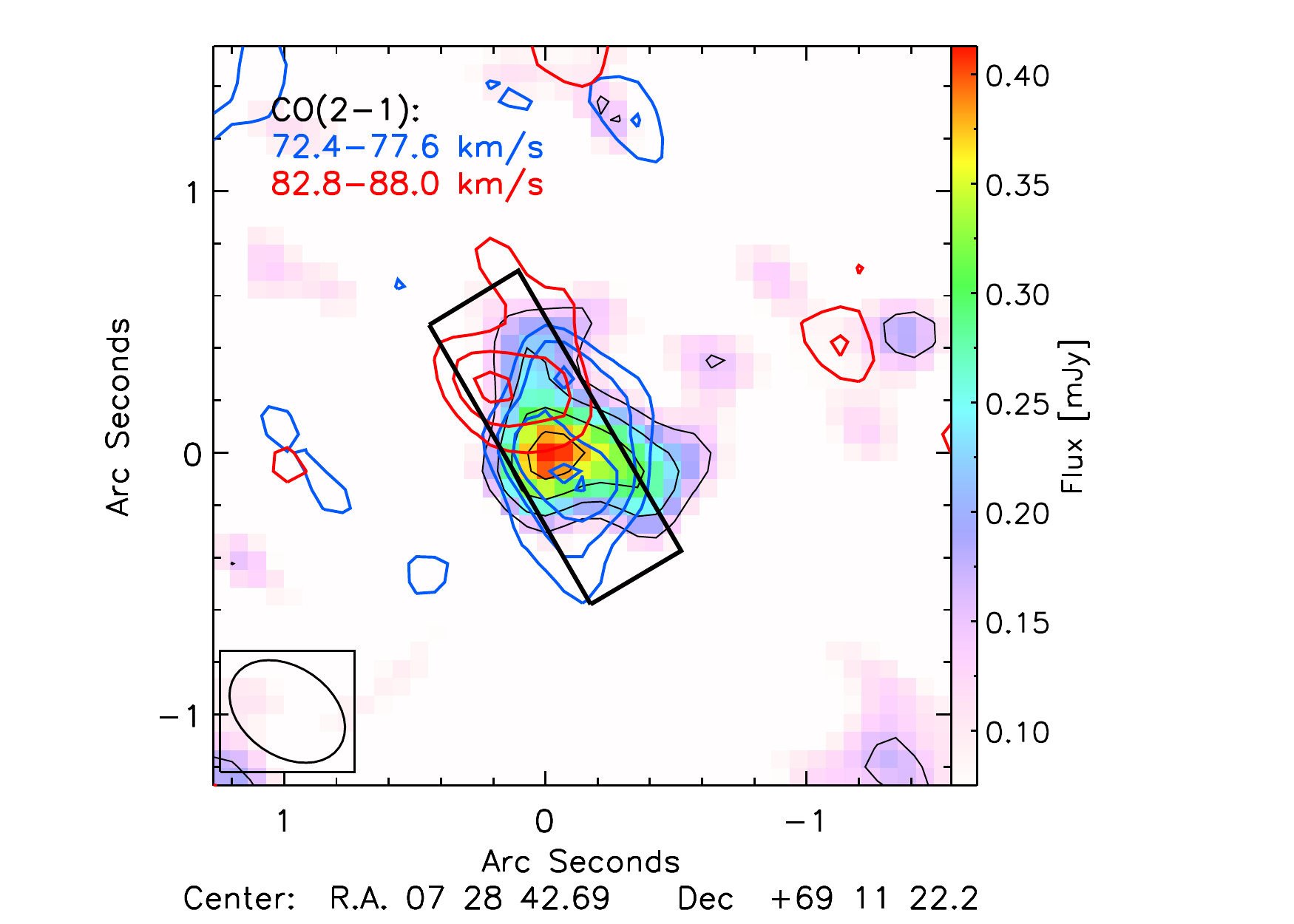}
  \caption{Panel $a$ (left) shows the CO(2--1) line profile for \mrka,
extracted from an area of 0$\farcs$68 size, defined as the emission
$>2\sigma$ in the map integrated over 72 -- 88 $\kms$.  The
two fitted gaussian components are shown with dashed blue and red
lines, and the single fitted gaussian with the solid green line.
Panel $b$ (right) shows the spatial distribution of the two components
in blue and red contours, corresponding to the respective kinematic
components in the left panel.  The continuum flux is indicated
by the color scale bar.  The contours for the continuum correspond to 2, 3, 4, and
5$\sigma$ with rms 0.075 mJy/beam.  The contours for the blue and red
components are integrated over 72.4 -- 77.6 $\kms$ and 82.8 -- 88.0
$\kms$, respectively, both showing contours starting at $2\sigma$ in
integer $\sigma$ intervals.  For the blue and red components, $1\sigma
= 12.9$ and 13.7 mJy/beam km/s, respectively.  The rectangle
indicates the region from which the PV diagram in
Figure~\ref{f_pvd} is calculated.  The NOEMA beam is also shown in the
lower left.
\label{f_components}}
\end{figure}

The observed CO(2--1) emission line is double-peaked, with a
velocity separation of 11.1 $\kms$ (Figure~\ref{f_components}a).
Figure~\ref{f_components}b shows the spatial distribution of the two
components, showing that they are not fully coincident.  At face value, the peaks are
offset from the continuum peak by about 1.7 pc (0.10$\arcsec$) and 6.0 pc
$(0.36\arcsec$) in projection, for the blue and red components,
respectively (Figure~\ref{f_components}b), although the beam size
  should be noted.  These may be separate, distinct clouds, with 7 pc
(0.42$\arcsec$) separation, in projection.  Both components appear to be
extended, although the red component is marginally resolved at the
3$\sigma$ level.  The blue component is resolved, and encompasses the
continuum peak in projection, allowing for the possibility that much
molecular gas is cospatial with the SSC stars.
The kinematics, line fluxes, and implied molecular masses of the two
components are given in Table~1, based on gaussian fitting
of the spectral lines.

To explain the presence of $10^5 \msol$ of molecular gas at such close
quarters to the young SSC, a first impression is that the gas could
simply be the cluster's natal material that is still in the process of being
destroyed through conventional mechanical and radiation feedback.
However, in high-density star-forming conditions, these feedback
processes are counteracted by radiative cooling and larger intra-cloud
gas pressure.  Our new  observations support this dynamic, also seen
in M82-A1 \citep{smith06}.

If the two CO components represent
an outflow, they could correspond to mechanical
feedback with an expansion radius and velocity of $R\sim 3.5$ pc and
$v\sim 5\ \kms$, respectively.  If the
observed molecular gas corresponds to material swept up within a
3.5-pc radius, the total molecular mass of $M_g \sim 10^5\ \msol$
corresponds to an original uniform density of $n_0\sim2\times 10^4\ \cc$, a
reasonable average value for the dense core of a massive, giant
molecular cloud.  At these high densities, cooling dictates that
either the wind cannot form \citep{silich17}, or the system
transitions rapidly from the energy-dominated to momentum-dominated
regime \citep[e.g., ][]{maclow88}.
The age, $t_6$ (Myr), of a momentum-driven shell can be 
estimated as $0.5 R/v = 0.3$ Myr.  The growth of the shell radius $R$ is
given by \citep[e.g.,][]{mccraysnow79},
\begin{equation}
  R = 16 \biggl(\frac{L_{36}} {n_0 v_{1000}}\biggr)^{1/4} t_6^{1/2}
  \quad \rm pc,
\end{equation}
where $L_{36}$ and $v_{1000}$ are the mechanical luminosity and
wind velocity in units of $10^{36}\ \ergs$ and $1000\ \kms$,
respectively, providing the impetus.  For $v_{1000} = 1$, appropriate
to massive star winds, this yields $L_{36} \sim 500$, a value
fully consistent with that expected for a young, $M_\star\sim 10^5\ \msol$ cluster at SMC
metallicity.  We further confirm that the observed dynamics are
inconsistent with a conventional, energy-driven bubble.  In this case, the
expansion velocity $v = 0.6 R/t$, implying an age of 0.4 Myr.  For an
adiabatic shell growth \citep[e.g.,][]{mccraysnow79},
\begin{equation}
  R = 27 \biggl(\frac{L_{36}} {n_0}\biggr)^{1/5} t_6^{3/5}
  \quad \rm  pc,
\end{equation}
the same parameters imply $L_{36}\sim 10$, over an order of magnitude
too small.   Thus, {\it energy-driven feedback does not dominate,}
supporting a catastrophic cooling scenario 
\citep{maclow88,silich07,krumholz09}.
The importance of cooling is also confirmed by 3-D
hydrodynamic simulations \citep{krause14,yadav17}.
It is therefore likely that radiation dominates the SSC feedback
\citep[e.g.,][]{freyer03}, especially if its age is $\leq 1$ Myr.  The
extinction-corrected \Ha\ luminosity is $8.4\times
10^{39}\ \ergs$ \citep{micheva17}, which is about
$10\times$ larger than the total wind power inferred above.  Although
the bipolar morphology implies only partial shells, this does not
affect the basic mechanical feedback calculations; certainly
adiabatic, pressure-driven feedback is impossible with partial shells.
We will quantitatively examine the possible feedback mechanisms in
detail in a future work.

\begin{figure}[ht!]
\epsscale{0.5}
\plotone{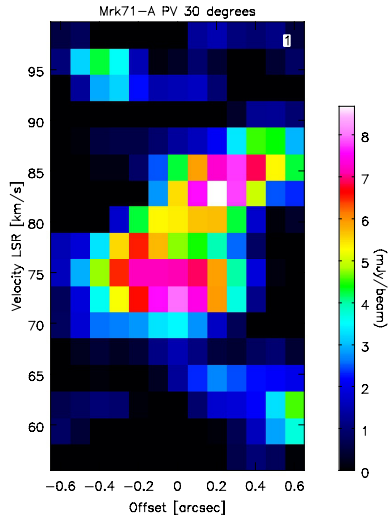}
 \vspace*{-1.0in}
\caption{Position-velocity diagram for CO(2--1) in the region of
  \mrka\ designated in Figure~\ref{f_components}b.  
\label{f_pvd}}
\end{figure}

Alternatively, the two molecular
components could be infalling, and perhaps be the remnants of two
molecular clouds whose collision triggered the formation of the
SSC.  This scenario has been suggested for other SSCs with similar CO
morphology by, e.g., \citet{fukui16}.
Figure~\ref{f_pvd} shows a position-velocity
(PV) diagram for \mrka, showing the projected kinematic structure.
\citet{haworth15} find from their simulated observations of
colliding clouds that the existence of ``broad bridge'' material between
the blue and red components is consistent with colliding
clouds.  The existence of emission between the blue and red components
in Figure~\ref{f_pvd} is morphologically similar to their models,
although our spatial resolution is relatively low.
The components could also represent random accretion
infall of vestigial material from the natal molecular environment.  

Taken as physically distinct clouds, their kinematics
are also fully consistent with a virialized system, together with the
SSC, as noted above.  The molecular gas is likely clumpy, perhaps dominated 
by the two unresolved, compact, massive clouds.  Since both peaks
are offset from the continuum peak (Figure~\ref{f_components}b), this
enhances the likelihood that the molecular clouds do not obstruct
the escape of ionizing radiation in our line of sight, consistent with the
low LyC optical depth suggested by \citet{micheva17}.
If the clouds are infalling, then the blue component is likely to be
behind the SSC, which further enhances this scenario.

Clarifying the timescale for gas retention and spatial relation to the SSC is
critical for understanding the conditions for LyC escape 
and Green Pea-like systems.  In particular, adiabatic superwinds have
been suggested to be important in clearing passages for 
ionizing radiation \citep[e.g.,][]{zastrow13,heckman11}.  The action
of mechanical feedback takes time, while after 3 Myr, the ionizing
stellar population declines \citep{dove00, fujita03}.  Thus,
in this model, there is significant LyC escape only in a short period
dominated by classical Wolf-Rayet stars around age 3 -- 5 Myr
\citep{zastrow13}.  However, \mrka\ is most likely younger than 3 Myr
old, perhaps even by a factor of 10, and it is a strong candidate
LCE, quantitatively matching Green Pea properties in all respects,
including extreme excitation and low optical depth \citep{micheva17}.
Since we show that any mechanical feedback in this system is not
energy-driven, it suggests that superwinds are not a
necessary condition for LyC escape.

\section{Conclusion}

In summary, our NOEMA CO(2--1) observations detect a compact, $\sim 7$ 
pc, $10^5\ \msol$ molecular cloud coincident with the SSC
\mrka, which is of similar mass.  At face value, the implied SFE
is high, on the order of 0.5, as seen in similar objects.  
In the extremely young, high-density star-forming conditions for
\mrka, energy-driven feedback will  
be suppressed by strong, radiative cooling \citep[e.g.,][]{silich07,
  krause14, yadav17}.
The presence of a massive, compact, molecular cloud
cospatial with the SSC is fully consistent with this expectation, and
we quantitatively demonstrate that any mechanical feedback from the
SSC must be momentum-driven.  Under these circumstances, 
radiation feedback from the young SSC is likely the dominant feedback
mode \citep[e.g.,][]{freyer03, krumholz09}.  Given that \mrka\ is an
extreme Green Pea analog and strong LCE candidate, our results suggest
that superwinds are not necessary to clear gas for LyC escape.

The CO(2--1) data appear to show two, spatially distinct, kinematic
components separated by $11\ \kms$.  If expanding, these could
be due to momentum-driven, stellar wind feedback.
Conversely, the components could be
colliding clouds responsible for triggering the formation of the SSC,
or simply random vestigial accretion.  Finally, the kinematics are also
consistent with a virialized system.

We also detect the nebular continuum in \mrka, allowing accurate
measurement of its absolute coordinates, and four additional
molecular clouds of similar masses within 50 pc of the SSC.  One is
unresolved and extremely compact, but no continuum is detected.

\acknowledgements
We thank Jim Dale, Mark Krumholz, Eric Pellegrini, Linda Smith, and
the anonymous referee for useful discussions.
This work is based on observations carried out under project number
W16BM with the IRAM NOEMA Interferometer.  IRAM is supported by
INSU/CNRS (France), MPG (Germany) and IGN (Spain).  

\vspace{5mm}
\facilities{NOEMA, HST(WFC3)}

\end{document}